\documentclass{article}

\title{Baryon Asymmetry of the Observed Universe as a Clue to a Resolution
 of  Dark Matter, Galaxy Formation and Other Standard Model Problems}
\author{A. Vankov}
\date{}

\begin{document}
\maketitle

\subsection*{Introduction}
          
      Antimatter is apparently present in the observed Universe in small quantities while physics
 shows no preference of matter over antimatter. The only way to save baryon symmetry and baryon charge
 conservation in cosmology is to suggest that the observed Universe is not the whole unique Universe
 but rather a huge matter-made fluctuation, or a representative of multitude of typical universes
 evenly made of either matter or antimatter and chaotically dispersed in infinite flat 3-space.
 In this approach all basic problems of the Standard Cosmological Model may be resolved in principal.
 As is known, the most significant of them, in addition to baryon symmetry problem, are next:

\noindent -  singularity (makes a cosmological theory incomplete)

\noindent -  flatness and horizon problems (as a consequence of singularity),

\noindent -  matter evolution (dark matter, galaxy formation and structuring, in particular),

\noindent -  CBR missing energy,

\noindent -  high energy tail in gamma and cosmic rays,

\noindent -  physical nature of sources of quasars and gamma bursts,

\noindent -  redshift distribution.

      It seems a little room is left to resolve them by simply refining the Standard Model. One can not
 suggest a remedy for, say, dark matter and galaxy formation problems without a revision of the whole
 chain, hence, coming at this to alternative cosmology.

      This work is devoted to a concept of Grand Universe (GU concept). The term means an infinite
 multitude of matter and antimatter typical universes as mentioned above. The work is an attempt to
 suggest the idea of an alternative cosmology resolving the basic Standard Model problems in principal.

\section{The observed Universe is limited in volume}

      Curiously enough, the idea of the universe being limited in volume is hidden in original
 Friedmann's solution for an expanding universe. The solution was found under asymptotic boundary
 condition, requiring empty space at however big distances to eliminate a set of free parameters in
 general solution of the Einstein's field equations. In fact, the observed Universe was never proved
 to be unlimited (if open). Just contrarily, some new evidences appeared that it has physical boundary.
 If so, the observed (better to say, home) Universe must have a center of expansion, an axis of rotation
 and  subsequently some asymmetric features due to our location somehow shifted off the center.
 Corresponding effects might be small enough, and higher precision of perspective  "beyond edge"
 and local observations is needed for their unambiguous treatment (global anisotropic effects in CBR
 and cosmic ray angular distributions, redshift and gamma burst counts, light and microwave polarization
 effects and others). Projects of search for antimatter are expected to be very informative.
 We should note that the purpose of this work is a discussion of consequences of the suggested concept,
 no matter how strong observational evidence is presently available in its support. However, there is
 one important fact that we want to pay a special attention to. It is an existence of absolute reference
 system (receding galaxies), or absolute aether (CBR). If the observed Universe is unlimited in space,
 this fact comes into conflict with special and general relativity theories and Newtonian physics as
 well because it breaks physical equality of inertial systems. It is a different situation when one
 considers a gravitationally linked system, which is limited in space by definition.

\section{The GU concept}

      In an infinite flat GU space an absolute reference system (physically distinguished material
 medium like absolute "ether") does not exist. It means that all material physical properties of GU
 matter on a big average are Lorentz invariant. One property already discussed is a baryon symmetry.
 The GU Universe is an even mixture of matter and antimatter, and the question arises, why both exist
 in separate forms not annihilating with each other. The answer comes from another GU matter property
 of being in a steady state of relativistic motion. This is the most difficult part of a description
 of the GU concept. A generalized transport theory must be developed allowing carrying out a computer
 simulation of a GU matter behavior in infinite space. However, it is quite understandable that its
 behavior must characterize a chaotic motion with a trend to form gravitationally linked systems.
 Chaos grows with growing a volume sample. It should be described by asymptotic uniform coordinate-momentum
 distribution, which is Lorentz invariant. In practical simulation a brightness parameter is expected to
 fluctuate in a range relevant to relativistic motion. One may conclude that the GU Universe must exist
 in a self-sustained state with an exact balance of matter and antimatter in bulk due to annihilation
 going in parallel with relativistic pair creation. In other words, the GU Universe lives a life in a
 way of constant cyclic recreation. It is again a subject of a computer simulation to follow physical
 processes of origination, evolution and decay of a typical universe as a part of a GU steady state
 evolution.

      If we accept our Home Universe having a physical boundary, the conclusion is that there is an outer
 "beyond edge" space filled with a relativistic baryon symmetric mixture in a form of all kinds of radiation,
 dust, clouds and different material objects (gravitationally linked smaller systems as well). They are
 remnants of decayed typical universes and play a role of constructing material in eternal recycling
 process of a recreation of universes of following generation. An interaction of a typical universe with
 what we may call a GU Background is an important factor in a typical universe evolution.

\section{Our Universe evolution}

      Suppose, we have developed a generalized GU matter transport theory. Its solution must give us a
 proper function, which is a mass distribution of gravitationally linked systems forming in a routine
 random process of interaction of the systems with a GU Background, each with other, in particular.
 It is easy to realize that the process provides for a growth of the systems but only a few of them
 among billions happen to reach a mature age of what we called typical universe. At a critical stage
 when gravitational forces can hardly hold all parts together a system becomes vulnerable. A criterion
 for starting decay is a ratio of potential to kinetic energy (or mass-to-radius ratio).

      Let us see what might have happened with our Home Universe 10-15 billion years ago. According to
 above criterion it should be in a stationary state with high chances to decay. The most probable
 triggering mechanism is a sudden mass drop due to a partial annihilation in a process of a random
 collision with an antimatter system of smaller mass. In this scenario the observed expanding Universe
 is a typical universe started decaying at some stationary critical state. It continues to decay until
 dissolving in GU Background. One must realize that in pre-expansion stage our Universe was a
 relativistic system as any other mature or embryonic formation in the GU Universe. A typical
 universe while evolving captures bodies from GU Background with any velocity below critical value.
 Our Home Universe before decay started was able to entrap bodies with fairly relativistic velocities.
 When the system has fallen apart its parts naturally aligned in Hubble's flow picture.

\section{Galaxy formation and dark matter issues}

      The Standard Model is quite successful in describing many processes going during expansion.
 In our opinion, everything concerning a "fireball" beginning is wrong and leads the Standard Model
 to deadlock. In a framework of suggested GU concept the Standard Model problems seem understandable.

      In above scenario of Home Universe evolution the Universe existed long time at a stationary
 pre-expansion state. It was a relativistic, more condensed than now mixture of mostly cold barionic
 matter (dust, rocks, dead stars etc). After first generation of stars had appeared a conversion of
 gravitational energy into heat began that made impact on further evolution. A considerable part of
 matter became hot and luminous. A system continued capturing matter and antimatter material from the
 GU Background, therefore, accumulating more and more internal energy, heating cold matter and enhancing
 star formation process. This stage had to be the right one for galaxy and galaxy cluster formation.
 Over time clustering process had to be more distinct. After the Universe started decaying the picture
 of evolution must be gradually changing in a sense that all processes are to slow down.

      Cold baryonic matter is a necessary component of typical universe medium. At the pre-expansion
 state it was not so much cold as at present. Now it is cooling as our Universe is expanding. We assume
 it being so called dark matter, which is distributed more or less uniformly with increased concentration
 within and around galaxies. It plays a role of local CBR sources. Hence, a CBR temperature is correlated
 with that of cold matter and is governed by a balance of absorbed and emitted energy with no missing.
 It is not surprising that cold matter is dominating but hardly detectable.

\section{Other physical issues in the GU concept}

     Coming back to the list of Standard Model problems we see the three last ones left without comments.
 Remember that a typical universe is exposed to super-high energy radiation coming from the GU Background.
 In the suggested concept this radiation plays a role of primer source of observed high energy tail in gamma
 and cosmic rays. As for quasars and gamma bursts these phenomena seem to have common physical nature, that
 is annihilation. The difference is that quasars are traced back to the "catastrophic" event trigged the
 decay (collision with antimatter universe) when huge mass formations are involved. Gamma bursts manifest
 "ordinary" collisions of matter and antimatter bodies of different masses. We assume that a lot of
 antimatter bodies are present in cold matter.

      The last issue concerns a test for a redshift distribution. The Standard Model failed in this test.
 Our study showed that the GU concept gives a full explanation of the problem.

\section{Conclusion}

      The suggested GU concept shows a way of resolving basic Standard Model problems at the expense of
 a radical revision of the Model. Many predictions may be drawn from the new concept. The concept is
 falsifiable but quantitative tests are needed on a base of a generalized transport theory to be developed.

\end{document}